\documentclass[12pt]{article}
\usepackage{graphicx}
\newcommand{\beq}{\begin{equation}}
\newcommand{\eeq}{\end{equation}}
\newcommand{\beqarray}{\begin{eqnarray}}
\newcommand{\eeqarray}{\end{eqnarray}}
\def\lsim{\raise0.3ex\hbox{$\;<$\kern-0.75em\raise-1.1ex\hbox{$\sim\;$}}}
\def\gsim{\raise0.3ex\hbox{$\;>$\kern-0.75em\raise-1.1ex\hbox{$\sim\;$}}}

\def\para{\vspace{0.3cm}\noindent}

\def\ev{\,{\rm eV}}

\def\mev{\,{\rm MeV}}
\def\gev{\,{\rm GeV}}

\begin{document}
\begin{flushright}
IMSc/2004/03/14
\end{flushright}
\begin{center}
{\large \bf On The Injection Spectrum of Ultrahigh Energy Cosmic 
Rays in the Top-Down Scenario\\}

\medskip

{Rahul Basu\footnote{rahul@imsc.res.in}}\\
{\it The Institute of Mathematical Sciences, Chennai 600 113,
INDIA.}

\bigskip

{Pijushpani Bhattacharjee\footnote{pijush@iiap.res.in}}\\
{\it Indian Institute of Astrophysics, Bangalore 560 034,
INDIA.}

\end{center}

\begin{abstract}
\noindent 
We analyze the uncertainties involved in obtaining the 
injection spectra of UHECR particles in the top-down scenario of their 
origin. We show that the DGLAP $Q^2$ evolution of  fragmentation functions
(FF) to $Q=M_X$ (mass of the X particle) from their initial values at low 
$Q$ is subject to considerable uncertainties. We therefore argue that, for 
$x\lsim 0.1$ (the $x$ region of interest for most large $M_X$ values of 
interest, $x\equiv 2E/M_X$ being the scaled energy variable), the FF 
obtained from DGLAP evolution is no more reliable than that 
provided, for example, by a simple Gaussian form (in the variable 
$\ln(1/x)$) obtained under the coherent branching approach to parton 
shower development process to lowest order in perturbative QCD.  
Additionally,  
we find that for $x\gsim0.1$, the evolution in $Q^2$ of the singlet FF, 
which determines the injection spectrum, is ``minimal'' --- the singlet FF 
changes by barely a factor of 2 after evolving it over $\sim$ 14 orders of 
magnitude in $Q\sim M_X$. We, therefore, argue that as long as the 
measurement of the UHECR spectrum above $\sim10^{20}\ev$ is going to 
remain uncertain by a factor of 2 or larger, it is good enough for most 
practical purposes to directly use any one of the available initial 
parametrisations of the FFs in the $x$ region $x\gsim0.1$ based on low 
energy data, without evolving them to the requisite $Q^2$ value. 
\end{abstract}
\newpage
\section{Introduction}
One of the main problems in understanding the origin of the 
observed Ultra-High Energy Cosmic Ray (UHECR) events with energy 
$E\gsim10^{20}\ev$\cite{uhecr_obs} --- below we will sometimes refer to 
these as Extreme Energy Cosmic Ray (EECR) events --- is the difficulty of 
producing such enormously energetic particles in astrophysical 
environments by means of known acceleration mechanisms. There are but a 
few astrophysical objects --- among which are, perhaps, 
Gamma Ray Burst (GRB) sources and a class of powerful radio galaxies --- 
where protons can in principle be accelerated to requisite energies (at 
source) of $\gsim10^{21}\ev$ by the standard diffusive shock acceleration 
mechanism albeit with optimistic assumptions on the values of the relevant 
parameters. However, even for these objects, their locations and spatial 
distributions are not easy to reconcile with the observed spectrum and 
large-scale isotropy of the UHECR particles. (For recent  
reviews on astrophysical source origin of EECR see, for example, 
Refs.~\cite{springer_book,torres_rev}).

\para 
An alternative mechanism of producing the EECR particles is provided by 
the so-called ``top-down'' (TD) scenario (see \cite{physrep} for a review) 
in which the EECR particles are envisaged to result from {\it decay} of 
some sufficiently massive particles, generically called ``X'' particles, 
of mass $M_X\gg10^{20}\ev$, which could originate from processes in the 
early Universe. This is in contrast to the conventional 
``bottom-up'' scenario in which {\it all} cosmic ray particles including 
the EECRs are thought to be produced through processes that accelerate 
particles from low energies to the requisite high energies in suitable 
astrophysical environments. 

\para 
The X particles of the TD scenario, if at all they exist in 
Nature, are most likely to be associated with some kind of new physics at 
some sufficiently high energy scale that could have been realized in an 
appropriately early stage of the Universe. 
Two possibilities for the origin of the X particles have been discussed in 
the literature: They could be short-lived particles released in 
the Universe today from cosmic topological defects such as cosmic 
strings, magnetic monopoles, etc.~\cite{td_book} formed in a 
symmetry-breaking phase transition in the early Universe. Alternatively, 
they could be some metastable (and currently decaying) particle species 
with lifetime larger than or of the order of the age of the Universe. 

\para 
Since the mass scale $M_X$ of the hypothesized X particle is well above 
the energy scale currently available in accelerators, its primary decay 
modes are unknown and likely to involve elementary particles and 
interactions that belong to unknown physics beyond the Standard Model 
(SM). However, irrespective of the primary decay products of the X 
particle, the observed UHECR particles must eventually result largely from 
``fragmentation'' of the Standard Model quarks and gluons,   
that come from the primary decay products of the X particles, into 
hadrons. The most abundant final observable particle species in the TD 
scenario are expected to be photons and neutrinos from the decay of the 
neutral and charged pions, respectively, created in the parton 
fragmentation process, together with a few percent baryons (nucleons). The 
injection- or the source spectra of various species of UHECR particles 
(nucleons, photons and neutrinos) in this TD scenario are thus ultimately 
determined by the physics of the parton fragmentation process. 
The final observable UHECR particle spectra are determined by further 
processing of these injection spectra due to extragalactic and/or 
Galactic propagation effects depending on where the X particle decay takes 
place. Clearly, in order to test the predictions of the TD scenario 
against UHECR experimental data, it is crucial to be able to reliably 
calculate the injection spectra of various UHECR particles in this 
scenario. This is the subject we concern ourselves with in this paper. 

\para 
The problem at hand is 
essentially the same as determining the single-particle inclusive spectrum of 
hadrons produced, for instance, in the process $e^+e^-\to \gamma/Z\to 
q\bar{q}\to {\rm hadrons}$ (see, for example, \cite{ellis_book}). The 
primary quarks produced in the collision would in general not be on-shell 
and would have 
large time-like virtuality $Q\sim \sqrt{s}$, the center-of-mass energy of 
the process. Each quark would, therefore, reduce its  
virtuality by radiating a gluon, the latter in 
turn splitting into a $q\bar{q}$ pair or into two gluons, 
and so on. This process gives rise 
to a parton shower whereby at each stage a virtual parton splits into two 
other partons of reduced virtualities. This process of parton shower 
development is well-described by perturbative QCD until the 
virtuality reduces to $Q=Q_{\rm hadron}\sim 1\gev$ when 
non-perturbative effects come into play binding partons into colorless 
hadrons. In the end, the link between partons and hadrons is 
quantitatively described in terms of 
fragmentation functions (FFs) $D_a^h(x,Q)$, 
which give the probability that a parton $a$ produced with an initial 
virtuality $Q=\sqrt{s}$ produces the hadron $h$ carrying a fraction 
$x\equiv 2E/\sqrt{s}$ of the energy of $a$ ($E$ being the energy of the 
hadron)\footnote{At high energies $E$ of our interest throughout this 
paper we shall assume $E\simeq p$, the momentum of the particle.}. The 
final single particle inclusive spectrum of hadrons is 
given by a convolution of these FFs with the production probabilities of 
the primary partons (see next section). 

\para 
In the same way, the problem of 
determining the injection spectrum of UHECR particles from the decay of X 
particles essentially reduces to determining the FFs $D_a^h(x,M_X)$ for 
various hadron species $h$ (pions, nucleons) where $a$ represents the 
primary partons to which the X particle decays. (Actually, in our 
present case, we will be interested only in the so-called ``singlet'' FF 
corresponding to a sum over all partons $a$ as explained later).     

\para 
Clearly, the FFs themselves cannot be directly calculated from first 
principles entirely within perturbative QCD without extra 
assumptions about the nature of the non-perturbative process of formation 
of hadrons from partons. Several different approaches have been taken  
in the recent literature for evaluating the relevant FFs, which are discussed 
below. 

\para
In this paper, we critically examine one of the approaches of evaluating 
the relevant FFs, namely, the DGLAP evolution 
equation method~\cite{fodor, sarkar-toldra,barbot-drees,barbot_thesis,aloisio}, 
that has been widely used in recent calculations of the UHECR injection 
spectra in the TD scenario. We discuss the inherent uncertainties involved 
in this approach in calculating the relevant FFs over the ranges of $x$ 
and $M_X$ of interest. We also compare the FFs so obtained with those 
given by a simple analytical expression (given by a Gaussian in the 
variable $\ln(1/x)$ as discussed later) obtained within the context of an 
analytical approach, namely, the coherent branching formalism, to lowest 
order in perturbative QCD~\cite{ellis_book}, this analytical approach 
being valid only under ``small" $x$ and ``large" $Q$ approximation. 
We show that except for ``large" $x\gsim 0.1$, the uncertainties 
involved in obtaining the relevant FFs by numerical solution of the DGLAP 
evolution equation do not allow much significant advantage of using this 
numerical method over the simple analytical (but 
approximate) formula for FFs provided by the coherent branching approach. 
At the same time, we also 
find that, in the region $x\gsim 0.1$, the evolution (in $Q$) of the {\it 
singlet} FFs (which is what we are interested in) is very 
little --- the singlet FF changes by only a factor of 2 or so after 
evolving it over $\sim$ 14 orders of magnitude in $Q\sim 
M_X$. We explain the reason for this, and argue that, as long as the 
measurement of the EECR spectrum is going to remain uncertain within a 
factor of 2 or larger (which is likely to be the case in the foreseeable 
future), it is good enough for most practical purposes to directly use any 
one of the available parametrisations of the FFs in the $x$ region 
$x\gsim0.1$ based on low energy (say at the 
Z-pole) data from $e^+e^-\to {\rm hadrons}$ experiments even 
without evolving them in $Q$ by means of DGLAP evolution equation.   

\para
As mentioned above, the X particle decay process may involve 
particles and interactions belonging to possible new physics 
beyond SM. Most of the recent studies using DGLAP 
evolution equation method have been done    
in the context of a particular model of 
the possible new physics beyond SM, namely, the Minimal Supersymmetric 
Standard Model (MSSM). While these studies are certainly useful, there 
exists, however, no direct evidence yet of Supersymmetry in general and 
the MSSM in particular. Indeed, the 
unknown nature of the physics beyond SM introduces additional 
uncertainties in the whole 
problem over and above the intrinsic uncertainties associated with the 
DGLAP evolution method itself which is fundamentally based on standard 
QCD. In order to analyze these uncertainties associated with 
the DGLAP evolution method itself, we restrict our analysis here to the 
standard DGLAP evolution equations for FFs based on QCD. 
Also, to keep our analysis simple, we shall illustrate our main results by 
considering the behavior of the FF for only one of the hadron species, 
namely, pions; our general conclusion, however, apply to nucleons as well 
as to other mesons like the K meson, too. 

\para
The rest of this paper is organized as follows: In the following section 
we set our notations and express the energy spectrum of hadrons resulting 
from the decay of the X particle in terms of the singlet fragmentation 
function (FF). In section 3, we review the various methods of evaluating 
the FF. Our main results are presented and discussed in section 4, and 
brief conclusions are presented in section 5. 

\section{Fragmentation Functions}
Let us consider the situation when the X decays 
from rest into a $q$ and a $\bar{q}$ pair (where $q$ can be 
$u,d,s,c,b,t$) which subsequently hadronize: 
$X\to q\bar{q}\to h + \cdots$ (here $h$ is a hadron). This is to 
facilitate direct comparison (at low c.~m. energies of 
$\sqrt{s}\sim100\gev$) with the available data on 
the similar process $e^+e^-\to \gamma/Z\to q\bar{q}\to h + \cdots$. 
We are interested in the energy spectrum or the single-particle energy 
distribution of the hadron species $h$, $dN^h/dx$, 
where $x\equiv 2E_h/M_X\leq 1$ is the scaled hadron energy. This can be 
written 
as a sum of contributions from different primary quarks $a=u,d,\ldots$ 
(and their antiparticles) as~\cite{ellis_book}
\beq
\frac{dN^h}{dx}(x,s)\propto \sum_{a}\int_{x}^{1}\frac{dz}{z} 
\frac{d\Gamma_{X\to a}}{dz}(z,s) D_a^h(x/z,s)\,,\label{F_h_1}
\eeq
where $d\Gamma_{X\to a}/dz$, the decay width of the X 
into parton $a$, is calculable in perturbation theory, and $D_a^h$ is the 
perturbatively non-calculable parton-to-hadron fragmentation function (FF). 

\para
Since the mass scale $M_X$ is much larger than the electroweak scale, 
we shall assume, following earlier work~\cite{sarkar-toldra}, flavor 
universality in the decay of X, which means that all primary quark 
flavors are produced with equal probability. This, together with the fact 
that, to lowest order for a 2-body decay, 
$d\Gamma_{X\to a}/dz\propto\delta(1-z)$, gives  

\beq
\frac{dN^h}{dx}(x,s)\propto \sum_{a}D_a^h(x,s)\equiv D_S^h\,,\label{F_h_2}
\eeq
where $D_S^h$ is the {\it singlet} FF \cite{ellis_book}. 

\para
The proportionality constant of equation (\ref{F_h_2}) can be 
determined from the energy conservation condition for hadronization of 
each individual quark, namely, 
\beq
\sum_{h}\int_{0}^{1}dx\, x D_a^h(x,s)=1\,,\label{D_energy_cons}
\eeq
together with the condition for overall energy conservation in the entire 
hadronization process, i.e., 
\beq
\sum_{h}\int_{0}^{1}dx\, x \frac{dN^h}{dx}(x,s)=2\,.\label{F_energy_cons}
\eeq
This finally gives
\beq
\frac{dN^h}{dx}(x,s)=\frac{1}{n_F}D_S^h(x,s)\,,\label{F_h_3}
\eeq
where $n_F$ is the number of active quark flavors.    

\section{Evaluation of FFs}
Three approaches to the problem of evaluating the relevant FFs have been 
followed in the literature. Below we discuss these in turn: 

\subsection{Using DGLAP evolution equation for FFs}
Although the FFs themselves are not 
directly calculable entirely within perturbative QCD, given their $x$ 
dependence extracted from experimental 
data at some scale $Q_0^2$, the {\it evolution} of the FFs with $Q^2$ is  
computable within perturbative QCD, and is given by the  
DGLAP evolution equation for FFs~\cite{ellis_book}. The relevant 
FFs at the scale $Q=M_X$ can then be evaluated by numerically solving the 
DGLAP evolution equation for the FFs, starting with input FFs extracted 
from $e^+e^-$ data at some laboratory energy scale, e.g., on the Z-pole 
($Q_0=91\gev$). This method has been used, for example, in 
Refs.~\cite{sarkar-toldra,barbot-drees,barbot_thesis,aloisio} to obtain 
the injection spectra of UHECR particles in the TD scenario. 

\subsubsection{Numerical solution of DGLAP evolution equation for FFs} 
The DGLAP evolution equation for the FF is given by a form similar to
that for parton distribution functions~\cite{ellis_book}
\begin{equation}
t\frac{\partial}{\partial t}D_i(x,t)=\sum_i\int_x^1 \frac{dz}{z}
\frac{\alpha_s}{2\pi}P_{ji}(z,\alpha_s)D_j(x/z,t),
\end{equation}
where the symbols have their usual meaning~\cite{ellis_book} and, as is 
well-known, the splitting function is $P_{ji}$ instead of
$P_{ij}$. These splitting functions have perturbative expansions 
in powers of the strong coupling $\alpha_s$ and we have taken the Lowest
Order (LO) expressions for these in our calculations of the  LO
DGLAP evolution for the FF. In practice, one considers non singlet
fragmentation combinations (in flavor space) of the form
$D_{NS}=D_{q_i}-D_{q_j}$ (where $i,j$ run over both quark and
anti-quark flavors) so that the flavor singlet gluons drop out, and the
singlet combinations $D_S=\sum_i D_{q_i}$ which mixes with the
fragmentation of the gluon, giving a matrix relation. Due to the $1/x$
pole in the $P_{gg}$ splitting function, the sea contribution increases
significantly at low $x$ for larger $Q^2$. In fact, the effect of
splitting is the same for distribution and fragmentation functions --- as
the scale of evolution $Q^2$ increases, the $x$ distribution is shifted
towards lower values. 

\para 
The evolution equations are usually solved numerically in Mellin space.
However, for convenience, we have used a numerical solution of these
equations in real space. 

\para 
There are various parametrisations for FFs available in the literature, 
given, for example, by KKP \cite{kkp}, BKK \cite{bkk}, and by Kretzer 
\cite{kretzer}, the most recent being those of KKP and Kretzer. These 
provide simple parametrisations of the FFs as functions of $x$ and $Q^2$ 
that are intended to reproduce their evolved values (obtained by solving 
the time-like evolution equations)
within the range of validity of their parametrisations. Most of these
parametrisations do not work below around $x\simeq 0.05$ or at the ultra 
high energy values of $Q^2$ that we are ultimately interested in.

\para 
Therefore, for numerical accuracy, we have not used any of the {\em
parametrisations} provided by these groups. 
We have taken the $x$ distributions of these FFs at 
their starting scale $Q_0^2$ and evolved them through the DGLAP 
equations 
to higher values of $Q^2$. This allows us to reach much higher 
values of $Q^2$ and very low values of $x \lsim 10^{-8}$ as is required 
for our analysis, way beyond the range of validity of the simple {\em
parametrisations} provided.
It is, of course, not clear whether even these starting values are 
reliable over such enormous ranges of $x$ and $Q^2$. For the present,
however, we will assume that
these starting parametrisations are reliable as long as we do not reach
ultra low values of $x$ where the phenomenon of coherent branching makes 
the FFs turn downwards as we go to lower $x$ (see below). 

\para 
In what follows, when we talk of a particular parametrisation (KKP, BKK
or Kretzer) it should be understood to mean that we use the initial
parametrisations provided by these groups and evolve them through the
evolution equations, and do not use the algebraic parametrisations given by the
authors valid over a restricted range of $Q^2$ and $x$.  

\subsection{Monte Carlo simulation}
In this approach one performs a direct numerical simulation of the 
parton shower process described by perturbative QCD coupled with a 
numerical modeling of the non-perturbative hadronization process. In the 
context of TD scenario of UHECR origin this ``Monte Carlo'' (MC) method 
has been studied in Refs.~\cite{ff_mc,aloisio}. A comparison of the DGLAP 
evolution and MC methods of obtaining the relevant FFs has recently been 
done in ~\cite{aloisio}. 

\subsection{Coherent Branching, Modified Leading-Log Approximation and
Local Parton-Hadron Duality: An analytical approach}
This is essentially an analytical approach entirely within
perturbative QCD in which the parton-to-hadron singlet FFs are obtained
from an analytical solution, obtained {\it under large $\sqrt{s}$ and
small $x$ approximations}, of a modified form of the DGLAP evolution
equation
that describes the parton shower evolution process within the so-called
``coherent branching'' formalism~\cite{ellis_book}. The method
assumes perturbative QCD to be valid all the way down to a virtuality
of $\sim\Lambda_{\rm eff}$, an ``effective'' QCD scale of order few
hundred MeV, and essentially gives the perturbative gluon-to-gluon
fragmentation function which dominates all FFs at small $x$ .
The FFs to different hadrons are taken to be proportional to this
gluon-to-gluon FF with appropriate normalization constants determined from
$e^+e^-\to {\rm hadrons}$ data in accordance with the hypothesis of Local
Parton Hadron Duality~\cite{lphd} which, at a purely phenomenological
level, seems to describe the experimental data rather
well~\cite{ellis_book}. Although there is no ``proof'' of the
LPHD hypothesis at a fundamental theoretical level yet, the basis of the
LPHD hypothesis is that the actual hadronization process occurs at a low
virtuality scale of order of a
typical hadron mass independent of the energy of the cascade initiating
primary parton, and involves only low momentum transfers and
local color re-arrangement which do not drastically alter the
form of the momentum spectrum of the particles in the parton cascade
already determined by the ``hard'' (i.e., large momentum transfer)
perturbative QCD processes. Thus, the non-perturbative hadronization
effects are lumped together in an ``unimportant'' overall normalization
constant which can be determined phenomenologically.

\para
The modification of the DGLAP evolution equation referred to above
consists of ordering the basic parton splitting processes (that give rise
to parton shower development) according
to decreasing emission angles between the final-state partons rather than
their decreasing virtuality. This angular ordering is due to the
color coherence phenomenon which leads to
suppression of soft gluon emission, making the FFs turnover at small $x$
below a characteristic value $x_c\sim (0.1\gev/\sqrt{s})^{1/2}$ --- an
effect clearly seen in the experimental data~\cite{pdg}.

\para
To leading order, the solution of the above mentioned modified DGLAP
equation gives the following Gaussian form for the singlet FF, $D_S$  
(dropping the superscript $h$), in the
variable $\xi\equiv\ln(1/x)$~\cite{ellis_book}:
\beq
D_S(\xi)\equiv xD_S(x,s)\propto
\exp\left[-\frac{1}{2\sigma^2}\left(\xi-\xi_p\right)^2\right] \,,
\label{mlla_gauss}
\eeq
where the peak position $\xi_p=Y/2$, and  
$2\sigma^2=(bY^3/36N_c)^{1/2}$, with $Y\equiv \ln(Q/\Lambda_{\rm 
eff})= \ln(m_X/\Lambda_{\rm eff})$ and $b=(11N_c-2n_F)/3$, $N_c=3$ being 
the number of colors.  

\para
Including the
next-to-leading order corrections, calculated in an analytical framework
known as Modified Leading-Log Approximation (MLLA)\cite{mlla_rev}, yields
again a closed form analytical expression for FFs that, as functions of
the variable $\xi$, can be well approximated by a ``distorted
Gaussian''~\cite{mlla_rev} in terms of calculable higher moments of the
variable $\xi$. The above Gaussian expression is a good approximation 
to the full MLLA result for $\xi$ not too far away on either side from the 
peak position $\xi_p$. The peak position $\xi_p$ also defines for us what 
we mean by ``small'' $x$ approximation: The MLLA (and its Gaussian 
approximation) are expected to be valid for $x$ not too large compared to 
$x_c\simeq (\Lambda_{\rm eff}/Q)^{1/2}$. 

\para
Within the LPHD picture, there is no way of distinguishing
between various different species of hadrons, all of which would thus have
the same spectral shape. Phenomenologically, the
experimental data at laboratory energies can be fitted by using different
values of $\Lambda_{\rm
eff}$ for different species of particles depending on their masses. For
our consideration of particles at EECR energies, however, all particles
are extremely relativistic (and hence essentially massless), and all
hadron species have essentially the same spectral {\it shape} which,
will be relatively insensitive to the exact value of $\Lambda_{\rm eff}$
since $\sqrt{s}\sim M_X\gg\Lambda_{\rm eff}$.

\para
Below, we shall compare the singlet FF obtained within the coherent
branching formalism described above with that obtained from numerical
solution of the DGLAP evolution equation. Since we consider DGLAP
evolution for the singlet FF only to leading order (LO), to be consistent,
and for simplicity, we shall use the corresponding LO result, namely, the
Gaussian expression given by eq.~ (\ref{mlla_gauss}) instead of the full
MLLA result. The Gaussian approximation (which we shall 
refer to as ``MLLA-Gaussian'' hereafter) becomes an increasingly 
better approximation to the full MLLA result at increasingly higher 
$\sqrt{s}$.

\para
An important point to note here is that, at laboratory energies, MLLA
gives a very good fit to the data at essentially {\it ``all''} $x$ values 
(including ``large'' $x$) for which data exist~\cite{mlla_rev}, although 
the MLLA analytic result is based on small $x$ approximation. For example, 
for $\Lambda_{\rm eff}=200\mev$ (which value we shall assume throughout 
this paper for illustration of the relevant numbers) and 
$\sqrt{s}=91\gev$, we have $x_c\simeq 0.05$. However, as shown in Figure 1 
below, the simple Gaussian curve provides a very good fit to the 91 GeV 
data at least up to $x\simeq 0.3$ and reasonably good fit at even larger 
values of $x$. Since the width of the Gaussian, $\sigma$, increases with 
$\sqrt{s}$ (albeit only logarithmically), we may expect the MLLA 
(Gaussian) to provide, with increasing $\sqrt{s}$, increasingly better 
description of reality at increasingly larger values of $x$ beyond the 
corresponding $x_c$ values. 

\para
Actually, this fact --- that MLLA results provide good description of the
data even at relatively ``large'' $x$ although it was derived under small 
$x$ approximation --- was already noticed in
\cite{mlla_book,mlla_rev} where this agreement was termed as ``natural,
though accidental''. The technical reason for this ``coincidence'' was 
also explained there; we shall, however, not go into these technical 
aspects in this paper. 

\section{Results and Discussions}
As a test of our DGLAP evolution code we show in Figure 1a the
comparison of the results of DGLAP evolution of the singlet FF for pion 
($\pi^++\pi^-$) with experimental data at 91.2 GeV \cite{sld_data} for 
the three different initial parametrisation (KKP, BKK, Kretzer) of the FFs. 
And Figure 1b shows the corresponding $D(\xi)$ vs $\xi$ curves.

\para 
The calculations are in overall good agreement with the data, as expected.
For comparison, we also display the MLLA-Gaussian curve. As
mentioned in the last section, the MLLA-Gaussian fits the data at large 
$x$ reasonably well. In fact, the Gaussian provides a better description 
of the data than the DGLAP results even at moderately large $x\sim 0.5$. 
And, as expected, at small $x$ ($x\lsim 0.1$) (i.e., $\xi\gsim 2.3$), the 
DGLAP results fail rather badly whereas the Gaussian gives an excellent 
fit. The reason for this is clear: The phenomenon of coherent branching 
dominates the parton shower process at low $x$. The 
standard DGLAP evolution equation for FF does not take this phenomenon 
into account, and the resulting FFs obtained from numerical solution of 
the DGLAP evolution equation are, therefore, not expected to be valid for 
$x\lsim x_c\sim 0.05$ (for $\sqrt{s}=91.2\gev$). (Actually, as seen from 
the figures, the DGLAP already fails at an $x$ value somewhat larger than 
this value of $x_c$). 

\para 
In Figure 2 we show the results for the singlet $D(x)$ (for pions) at 
various values of $M_X$ up to $M_X=10^{16}\gev$ obtained by solving the 
DGLAP equation for three different initial FF parametrisations. Again, for 
comparison we also show the MLLA-Gaussian curves. 

\para 
In Figure 3 we show the $D(\xi)$ vs $\xi=\ln(1/x)$ curves for the same set of
parametrisations as in Figure 2.

\para 
In Figures 2 and 3, we have normalized the Gaussian curves with 
the DGLAP evolution results at $x\simeq 0.03$ where the results of all 
three FF parametrisations agree. 
It can be seen from Figures 2 and 3 that there are large discrepancies 
amongst the results of the three different initial parametrisations for 
$x\lsim10^{-2}$. 
Note that these discrepancies are at $x$ regions well above the turning
points of the FFs that are due to coherence effects, and are therefore to be
attributed to magnification (due to $Q^2$ evolution) of the intrinsic
differences amongst the three initial parametrisations. The
parametrisations are done by fitting the FFs to the known data which go
only up to $\sqrt{s}\sim$190 GeV. Moreover, most parametrisations
(including KKP) are restricted to $x$ region above $\sim 0.05$ (because
there are no data for lower $x$ at the initial scale of parametrisation).
So the resulting initial parametrisations do not satisfy the various sum
rules very well. For example, the momentum (or energy) sum rule is
rather poorly satisfied in KKP. Also, the behavior of $D(x,Q)$ shows
some strange behavior as illustrated more clearly in Figures 4 a--c where
we show
the behavior of FF as a function of $x$ for different values of $Q$ for
KKP, BKK and Kretzer parametrisations.

\para 
On standard theoretical ground, it is expected
that with increasing $Q$, the $x$ distribution should shift towards
lower values, i.e., the FF should increase with
$Q$ at low $x$ and decrease at large $x$. In effect, this implies a
steepening of the particle spectrum with increasing $Q$. Thus,
the FFs as a function of $x$ for two different values of $Q$ should cross
at some $x$. However, the curves in Figures 4a--c do not show this
expected crossing behavior except marginally for the BKK parametrisation 
(Figure 4a) at low $Q$ values (specifically the $Q=$10 and 90 GeV curves). 
This is a reflection of the fact that the
data available at existing energies (on which the parametrisations
are based) show this behavior clearly only at
low $Q$ ($\sqrt{s}<50\gev$), while being essentially flat beyond this
value for all $x$ (see, e.g., Figure 15.1(b) in Ref.~\cite{pdg}).
Moreover, none of the parametrisations use the low $x$ data which
do show slight increase with $Q$ (see Figure 15.1(b) in 
Ref.~\cite{pdg}). Consequently, our evolution results based on these 
parametrisations also do not show this effect. In fact, the 
$Q=10^{16}\gev$ curve is always substantially below the curves for lower 
$Q$ for all $x$ reflecting the above facts.

\para
The above results illustrate the fact that using DGLAP evolution to 
predict the shape of the UHECR injection spectra is subject to 
considerable uncertainty associated with the initial FF parametrisations. 

\para
The other important point to notice is that the effect of the evolution of 
the singlet FF with $Q^2$ is ``minimal''. In fact, over the
whole range of $M_X$ from $91$ -- $10^{16}\gev$, the FF changes only by 
a factor $\sim2$ (see Figure 2). The reasons for this
is that the $Q^2$ evolution of the FFs is driven mainly by the gluon.
However, in our case, particularly at very large $Q$ and small $x$, the
gluon FF is several orders of magnitude smaller than the singlet FF.
Therefore, the evolution of the gluon FF has very little effect on the
singlet FF. Actually, with the initial parametrisations used here, the
singlet is 4 orders of magnitude larger than the gluon even at smaller
$Q$ (2.5 GeV) for small $x$ ($\sim 10^{-7}$). Hence over the whole range
in $Q$ (i.e., up to $M_X\sim10^{16}\gev$), there is very little
evolution with $Q$.

\para
So it appears that full DGLAP evolution is
essentially unnecessary at the current level of measurement of the UHECR
spectra which are, and likely to remain in the foreseeable future,
uncertain by factors larger than 2 or so. A typical parametrisation of
the FFs is of the form $\sim x^\alpha (1-x)^\beta $ with $\alpha$ and
$\beta$ being functions of $Q^2$. However, the above discussion seems to
suggest that, as far as the singlet FF (for a given hadron species) is 
concerned, it is sufficient to obtain it directly from the individual FFs 
of different partons as given by the above form with appropriate values of 
the parameters $\alpha$ and $\beta$ extracted from the relevant 
experimental data at some laboratory energy scale $Q_0$.   

\para
To what extent can one use the MLLA at all $x$ values of 
interest, namely, in the region $x\lsim0.1$? While, as we have seen, the 
MLLA describes the data well essentially at all $x$ for 
$Q=91\gev$, the situation becomes more complicated for larger values of 
$\sqrt{s}=M_X$. For $M_X=10^{13}\gev$, for example, the coherent branching 
effect becomes important only at ``ultra-low'' $x\lsim x_c\sim 
1.4\times10^{-7}$. At the same time, the TD scenario of UHECR origin is 
generally relevant only for observed UHECR energies $E>10^{10}\gev$, which 
corresponds to $x>2\times10^{-3}\gg x_c$ for $M_X=10^{13}\gev$. Thus, the 
coherent branching effects are not yet ``switched on'', and  
it is not a priori clear whether the MLLA expression for the FF 
is valid at such relatively ``large'' $x$. This is the basis of the 
argument that one should not use the MLLA results in these 
circumstances; instead, one should obtain the relevant FFs by solving the 
DGLAP evolution equation for FF. While this is perhaps what one {\it 
should} do, the problem here is that the starting 
parametrisations of the FFs are not known at such values of $x$, and one 
has to extrapolate the starting FFs well below the lowest $x$ value ($\sim 
0.05$) up to which the initial parametrisations of the FFs are known. This 
extrapolation is fraught with considerable 
uncertainty since one has to {\it assume}, a priori, a form of the 
extrapolated FF, and, as discussed above, simple 
extrapolation of the existing FF parametrisations to small $x$ values 
gives widely different answers when evolved to high $M_X$ values by means 
of DGLAP evolution equation. 

\para 
In Ref.~\cite{barbot_thesis}, the guiding 
principles adopted for extrapolation of the starting FFs to the 
relevant low $x$ values are energy conservation and continuity of the FFs. 
These conditions, however, do not uniquely fix the form of the FFs valid 
over the entire range of $x$ of interest. In addition, to impose energy 
conservation, Ref.~\cite{barbot_thesis} had to assume FFs for all hadrons 
to 
have the same power-law form at low $x$, based on MLLA-LPHD result. The 
interesting result of Ref.~\cite{barbot_thesis}, however, is that the 
resulting FFs obtained by solving the DGLAP evolution equation at 
high $M_X$ smoothly match onto the properly normalized MLLA result 
at an $x$ value which is considerably larger than the corresponding value 
of $x_c$. This suggests that one might as well use the MLLA-LPHD 
formula in the region $x\lsim0.1$, considering the uncertainties involved 
in DGLAP evolution of FFs. 

\section{Summary and conclusions} 
In this paper we have analyzed the uncertainties involved in obtaining the 
injection spectra of UHECR particles in the top-down scenario of their 
origin. We have demonstrated that evaluating the relevant FFs at the 
values of $M_X$ and $x$ of interest by evolving them (in $Q=M_X$) from 
their initial (parametrised) values at low $Q$ by numerically solving 
the DGLAP evolution equation for FF is subject to considerable 
uncertainties. Indeed, we find that for $x\lsim 0.1$ (the $x$ region of 
interest for most large values of $M_X$ of interest), the FF obtained 
from DGLAP evolution cannot be said to be any more reliable than that 
provided by the simple Gaussian form (in the variable $\xi$) based on 
coherent branching approach to parton shower development. At the same 
time, we also find that for $x\gsim0.1$, the evolution of the singlet FF, 
which determines the injection spectrum, is ``minimal'' --- the singlet FF 
changes by barely a factor of 2 after evolving over $\sim$ 14 orders of 
magnitude in $Q\sim M_X$. We, therefore, argue that as long as the 
measurement of the EECR spectrum is going to remain uncertain by a 
factor of 2 or larger (which is likely to be the case in the foreseeable 
future), it is good enough for most practical purposes to directly use any 
one of the available {\em initial} parametrisations of the FFs in the $x$ 
region $x\gsim0.1$ based on low energy (say at the 
Z-pole) data from $e^+e^-\to {\rm hadrons}$ experiments, without any
need for evolving them to the required EECR $Q^2$ value.

\para
{\bf Acknowledgments}\\
This work was begun at the Seventh Workshop on High Energy Physics 
Phenomenology (WHEPP-7) held at Harish-Chandra Research Institute, 
Allahabad, India, January 4--15, 2002. We thank all the organizers and 
participants of that Workshop for providing a stimulating 
workshop environment. The work of PB is partially supported by a NSF 
US-India cooperative research grant. RB would like to thank D.~Indumathi
for useful discussions. 

\newpage
\begin{center}
\begin{figure}
\includegraphics[height=7cm]{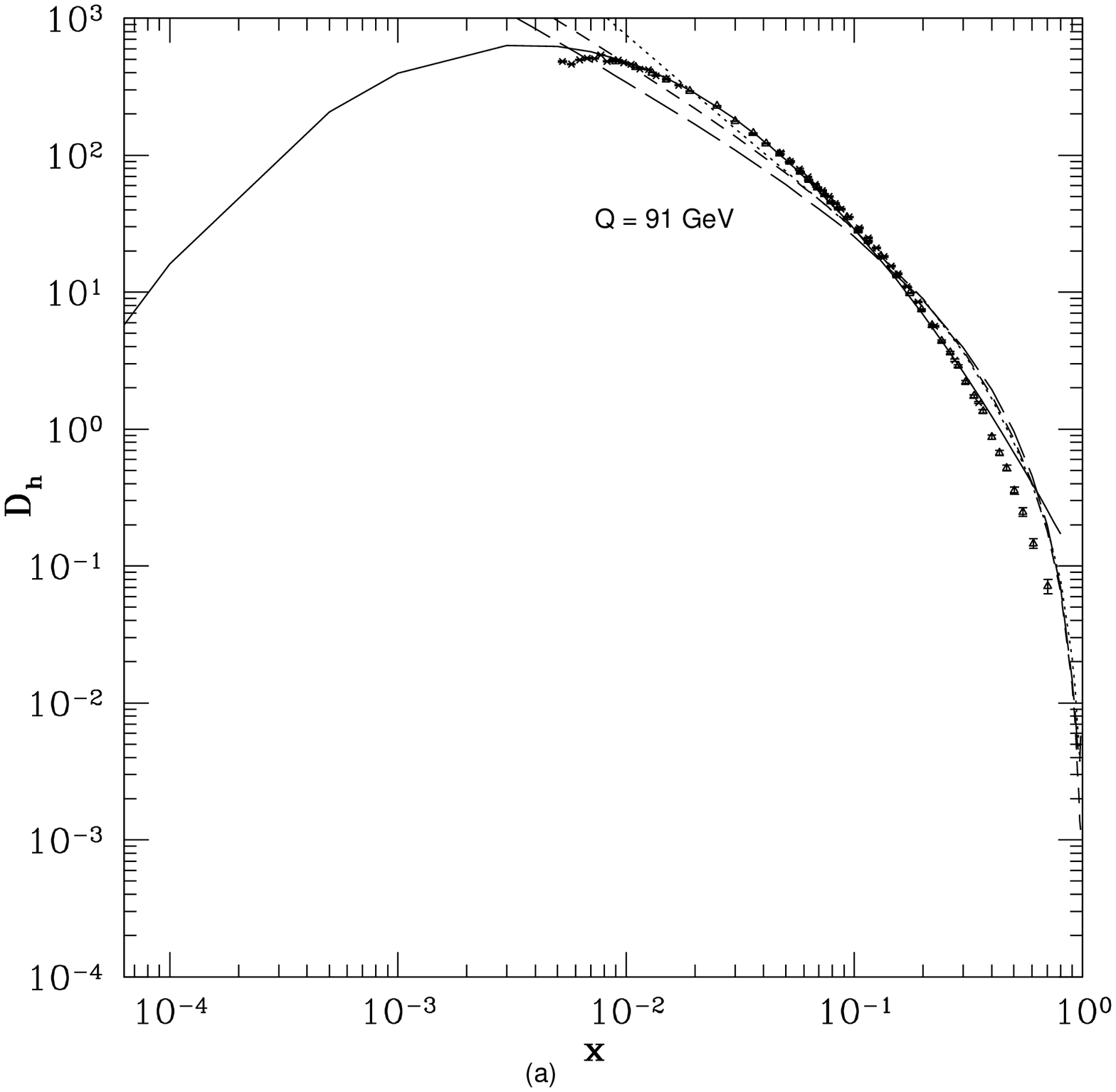}\hfill
\includegraphics[height=7cm]{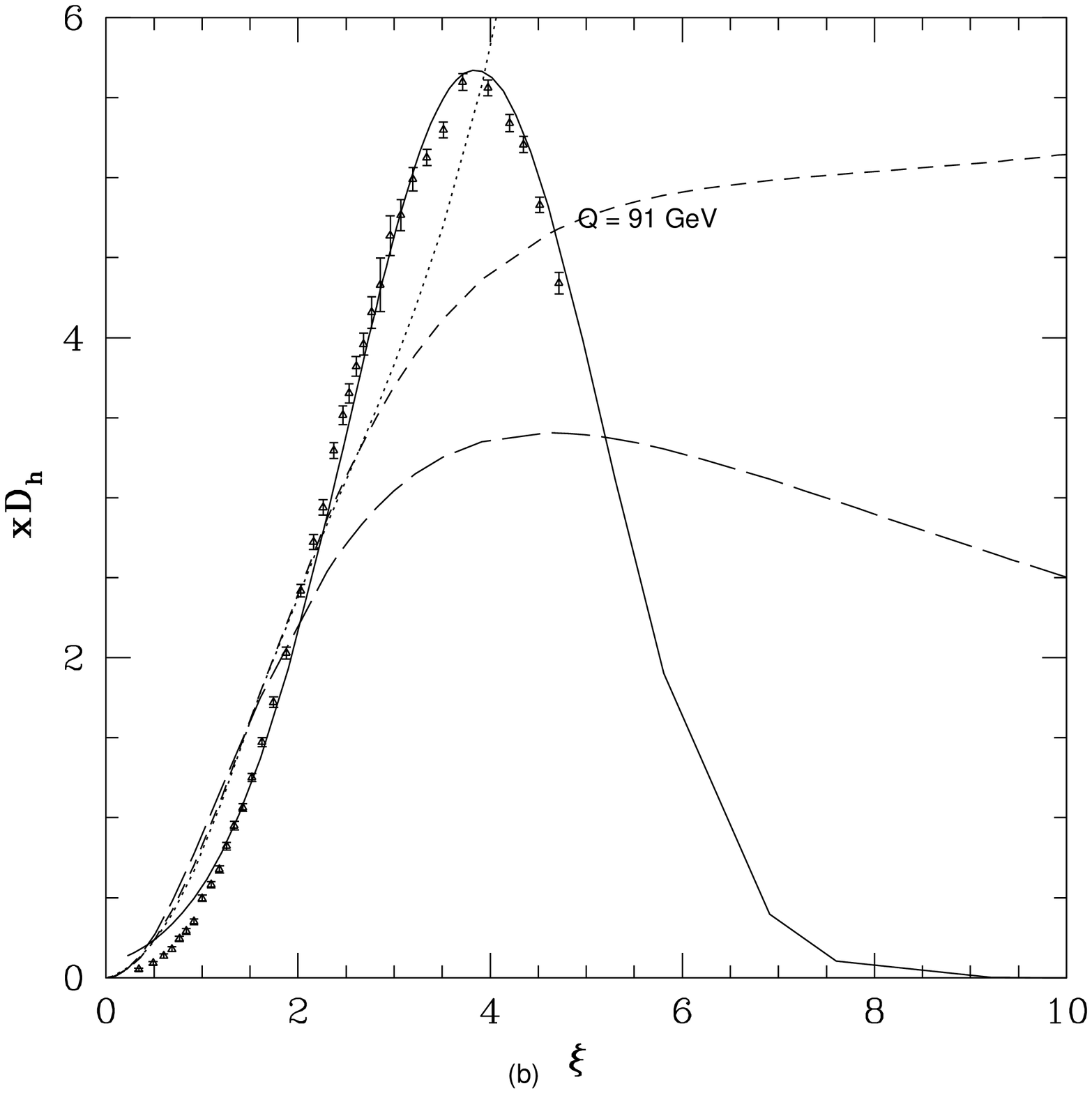}
\caption{$D(x)$ and $D(\xi)$ curves along with 91 GeV data for three
different parametrisations: KKP (dotted), BKK
(short-dash) and Kretzer (long-dash). Also shown is the MLLA-Gaussian
curve (solid line) given by equation (\ref{mlla_gauss}) with normalization
fixed by average pion multiplicity data.}
\end{figure}
\end{center}
\begin{center}
\begin{figure}
\includegraphics[scale=0.8]{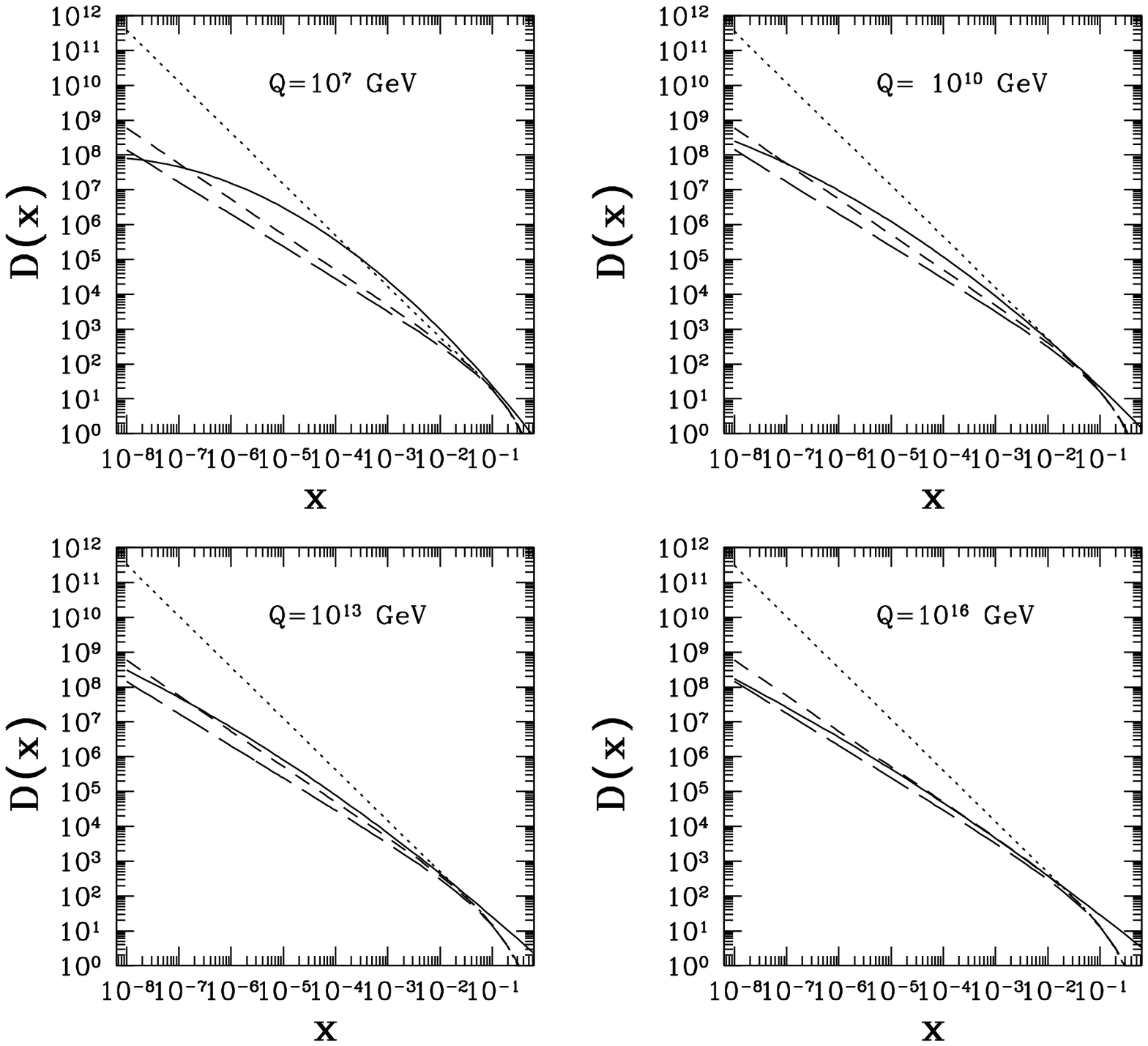}
\caption{A comparison of $D(x)$ vs $x$ curves at various different 
values of $Q=M_X$ for the three different FF parametrisations : KKP 
(dotted), BKK (short-dash) and Kretzer (long-dash). The solid curves 
represent the MLLA-Gaussian.} 
\end{figure}
\end{center}
\begin{center}
\begin{figure}
\includegraphics[scale=0.8]{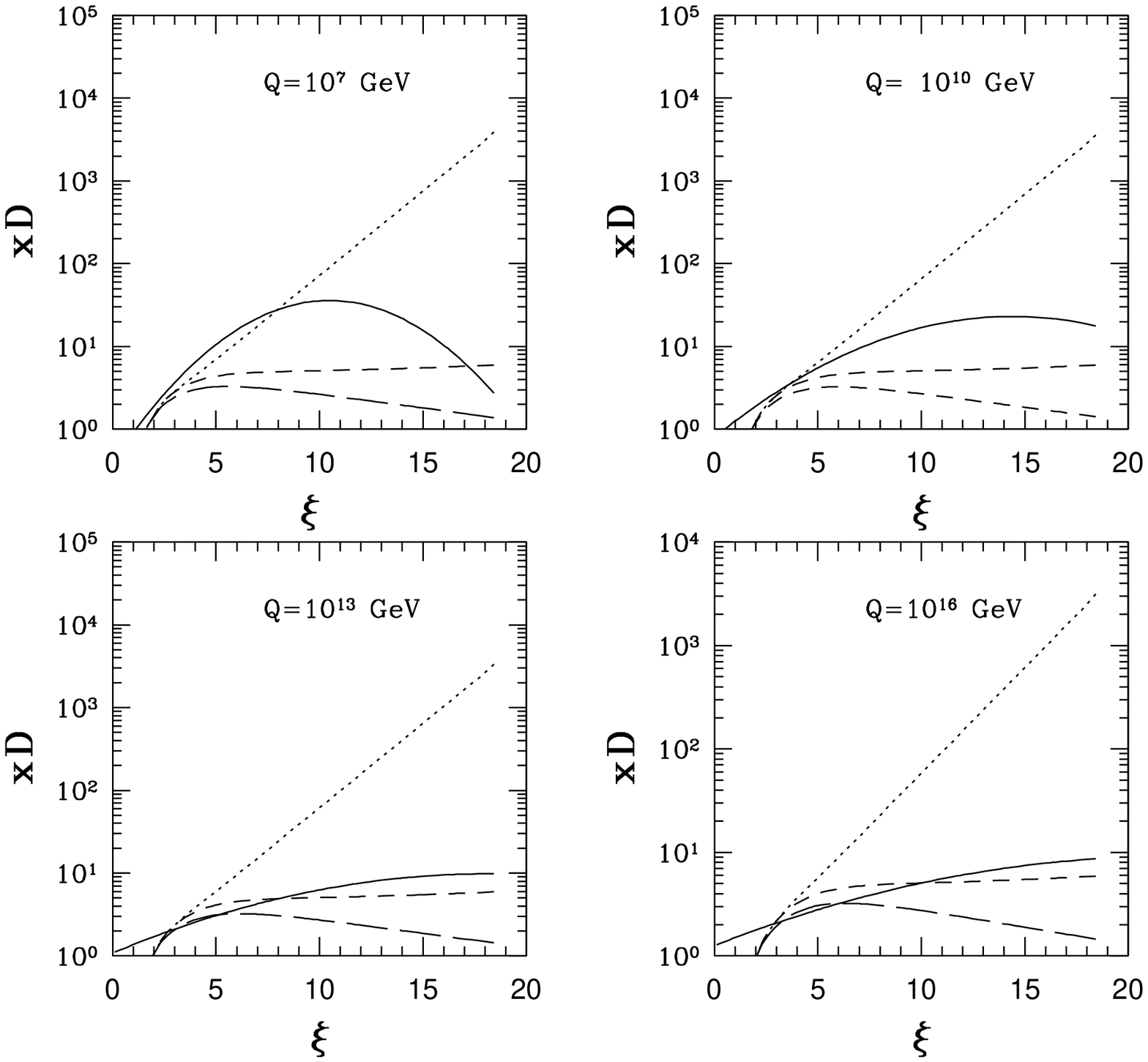}
\caption{A comparison of $D(\xi)$ vs $\xi$ curves at various different 
values of $Q=M_X$ for the three different FF parametrisations : KKP 
(dotted), BKK (short-dash) and Kretzer (long-dash). The solid curves 
represent the MLLA-Gaussian.} 
\end{figure}
\end{center}
\begin{center}
\begin{figure}
\includegraphics[scale=0.3]{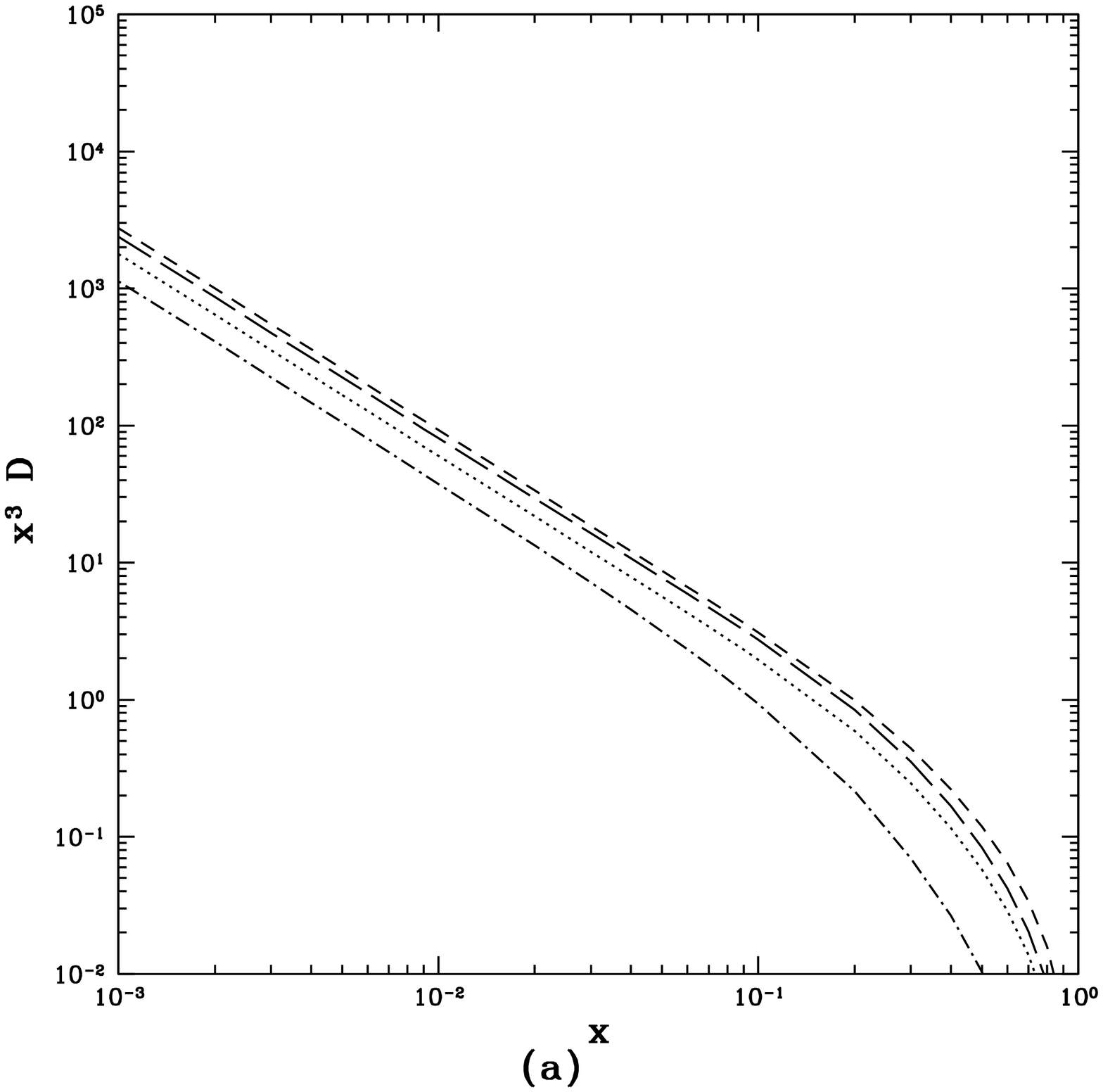}\hfill
\includegraphics[scale=0.3]{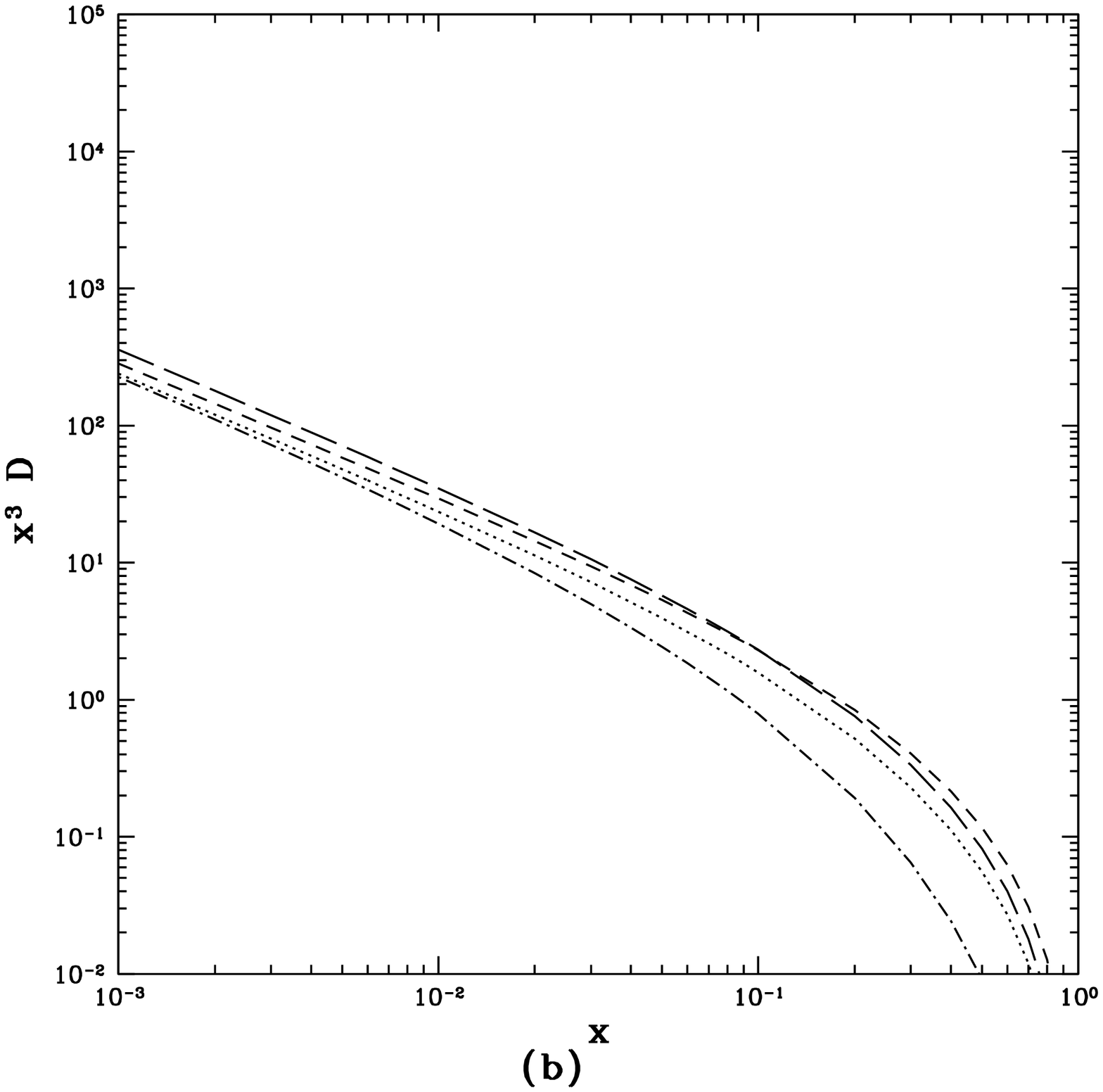}\\
\begin{center}
\includegraphics[scale=0.3]{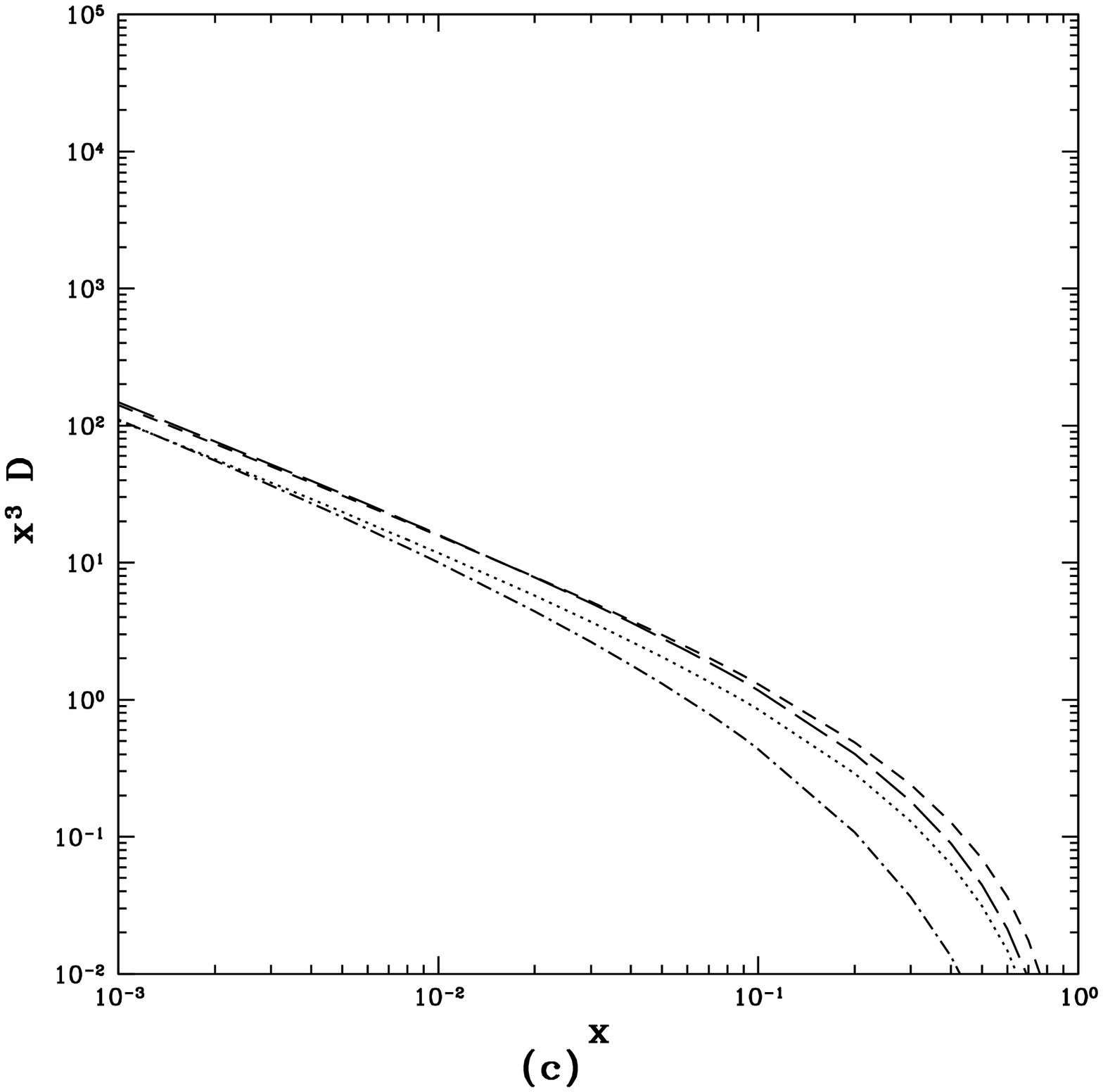}
\end{center}
\caption{A plot of $x^3 D(x,Q)$ vs. $x$ for (a) KKP (b) BKK and (c)
Kretzer FF parametrisations. For each of these, the lines correspond to 
$Q$=10 (short-dash), 91 (long-dash), 189 (dot) and $10^{16}$ (dot-short 
dash) GeV.} 
\end{figure}
\end{center}


\begin{thebibliography}{99}
\bibitem{uhecr_obs} M. Takeda et al (AGASA Collaboration), Astropart.
Phys. 19 (2003) 447; T. Abu-Zayyad et al (HiRes Collaboration), 
arXiv:astro-ph/0208243, astro-ph/0208301.

\bibitem{springer_book} M. Lemoine and G. Sigl (Eds.), {\it Physics and 
Astrophysics of Ultra-High-Energy Cosmic Rays} (Springer, Berlin, 2001).  

\bibitem{torres_rev} D.F. Torres, L.A. Anchordoqui, 
arXiv:astro-ph/0402371. 

\bibitem{physrep} P. Bhattacharjee and G. Sigl, Phys.~Rep. 327 (2000) 109. 

\bibitem{td_book} A.~Vilenkin and E.~P.~S.~Shellard, {\it Cosmic Strings
and other Topological Defects} (Cambridge Univ. Press, Cambridge, 1994).

\bibitem{ellis_book} R.K. Ellis, W.J. Stirling and B.~R.~Webber, {\it QCD 
and Collider Physics} (Cambridge Univ. Press, Cambridge, 1996). 

\bibitem{fodor}Z.~Fodor and S.~D.~Katz, Phys.\ Rev.\ Lett.\  {\bf 86}
(2001) 3224 arXiv:hep-ph/0008204

\bibitem{sarkar-toldra} S. Sarkar and R. Toldra, Nucl.~Phys. B621 (2002) 
495-520. 

\bibitem{barbot-drees} C. Barbot, M. Drees, Astropart.~Phys. 20 (2003) 5. 

\bibitem{barbot_thesis} C. Barbot, arXiv:hep-ph/0308028.  

\bibitem{aloisio} R.~Aloisio, V.~Berezinsky and M.~Kachelrie\ss, 
Phys.~Rev.~D (to appear) [arXiv:hep-ph/0307279]. 

\bibitem{kkp} B.~A.~Kniehl, G.~Kramer and B.~P\"otter, Nucl.~Phys. B582 
(2000) 514. 

\bibitem{bkk} J. Binnewies, B.~A. Kniehl and G.~Kramer, Phys.~Rev. D52 
(1995) 4947. 

\bibitem{kretzer} S. Kretzer, Phys. Rev. D62 (2000) 054001. 

\bibitem{ff_mc} M. Birkel and S. Sarkar, Astropart.~Phys. 9 (1998) 297; 
V.~S.~Berezinsky and M.~Kachelrie\ss, Phys.~Rev.~D63 (2001) 034007.  

\bibitem{lphd} Ya.~I.~Azimov, Yu.~L.~Dokshitzer, V.~A.~Khoze, and
S.~I.~Troyan, Z.~Phys. C 27 (1985) 65; C 31 (1986) 213.

\bibitem{pdg} K. Hagiwara et al [Particle Data Group], Phys. Rev. D66 
(2002) 010001. 

\bibitem{mlla_rev} For a review, see, for example, V.~A.~Khoze and 
W.~Ochs, Int. J. Mod. Phys. A12 (1997) 2949. 

\bibitem{mlla_book} Yu. L. Dokshitzer, V.~A.~Khoze, A.~H.~Mueller, and
S.~I.~Troyan, {\it Basics of perturbative QCD} (Editions Frontiers,
Saclay, 1991). 

\bibitem{sld_data} K. Abe et al [SLD Collaboration], Phys.~Rev.~D59 (1999) 
052001. 

\end{thebibliography}
\end{document}